\documentstyle[e-e-ijmpa,twoside]{article}
\input psfig
\input epsfig.sty
\def\nn{\noindent}
\def\ie{{\it i.e.}}
\def\eg{{\it e.g.}}

\def\etal{{\it et al.}}

\renewcommand{\thefootnote}{\fnsymbol{footnote}}	

\begin{document}


\rightline{\vbox{\halign{&#\hfil\cr
&SLAC-PUB-7675\cr
&October 1997\cr}}}
\vspace{0.5in}

\normalsize\textlineskip
\thispagestyle{empty}
\setcounter{page}{1}

\title{{LEPTOQUARKS AT FUTURE LEPTON COLLIDERS}
\footnote{To appear in the {\it Proceedings of the $2^{nd}$ International
Workshop on $e^-e^-$ Interactions at TeV Energies}, Santa Cruz, CA,
22-24 September 1997}
}

\author{ {THOMAS G. RIZZO}%
\footnote{Work supported by the Department of Energy, 
Contract DE-AC03-76SF00515}
}

\address{Stanford Linear Accelerator Center,\\
Stanford University, Stanford, CA 94309, USA\\
E-mail: rizzo@slacvx.slac.stanford.edu}

\maketitle\abstracts{In this talk I summarize the capability of future 
lepton colliders to discover leptoquarks and to determine their 
electroweak quantum numbers. This analysis is an updated discussion 
based on the results presented in the Snowmass 1996 New Phenomena Working 
Group report as well as some more recent work that has appeared in the 
literature as a result of the HERA high-$Q^2$ excess.}

\setcounter{footnote}{0}
\renewcommand{\thefootnote}{\alph{footnote}}

\vspace*{1pt}\textlineskip	

\section{Introduction and Background}

The observation of a possible excess of both neutral and charged current 
events at HERA{\cite {hera}} has brought the discussion of leptoquarks(LQs) 
and their properties off the back burner, though they have been subjects of 
study for quite some time{\cite {brw,rev}}. 
For more than a decade any discussion of leptoquark models has 
been historically based on the classic work by Buchm\" uller, R\" uckl and 
Wyler (BRW){\cite {brw}}.  As these authors showed, LQs can be either 
spin-0 (scalars) or spin-1 (vectors) and may carry fermion number, $F=3B+L=0$ 
or $\pm 2$. In that paper the authors provided a set of basic 
assumptions under which consistent leptoquark models can be constructed; these 
can be restated as:
\begin{tabbing}
(a) \= LQ couplings must be invariant with respect to the Standard Model (SM)\\
\> gauge interactions,\\
(b) \> LQ interactions must be renormalizable,\\
(c) \> LQs couple to only a single generation of SM fermions,\\
(d) \> LQ couplings to fermions are chiral,\\
(e) \> LQ couplings separately conserve Baryon and Lepton numbers,\\
(f) \> LQs only couple to the SM fermions and gauge bosons.
\end{tabbing}
Amongst these assumptions, both (a) and (b) are can be considered sacrosanct 
whereas (c)-(e) are data driven{\cite {brw,rev}} by a host of low energy 
processes as discussed by Davidson, Bailey and Campbell and by Leurer. 
Assumption (f) effectively requires that the leptoquark be the only 
new component added to the SM particle spectrum, which seems quite unlikely in 
any realistic extended model, and is perhaps the least tenable. 
Assuming the validity of (a)-(f), the possible set of LQ models is quite 
restricted and we arrive at the relatively short list shown in 
Table~\ref{lqprop} as obtained years ago by BRW:

{\small
\begin{table}
\tcaption{Quantum numbers and fermionic couplings of the BRW leptoquark 
states as well as the minimal $SU(5)$ representation into which it can be 
embedded.  No distinction is made between the $SU(5)$ representation and 
its conjugate.}
\begin{center}
\begin{tabular}{l|ccrccc}
\hline\hline
 & Leptoquark \rule{0pt}{15pt} 
 & SU(5) Rep & & $Q$  & Coupling & $B_\ell$ \\ \hline
Scalars & & & & & & \\ \hline
$F=-2$ \rule{0pt}{15pt}& $S_{1L}$        
       & {\bf 5} & 
       & $1/3$ & $\lambda_L\, (e^+\bar u)$, 
$\lambda_L\, (\bar\nu\bar d)$ & $1/2$ \\
       & $S_{1R}$        
       & {\bf 5} & 
       & $1/3$ & $\lambda_R\, (e^+\bar u)$ & 1 \\
       & $\widetilde S_{1R}$ 
       & {\bf 45} & 
       & $4/3$ & $\lambda_R\, (e^+\bar d)$ & 1 \\[2ex]
       &         &        
       & & $4/3$ & $-\sqrt 2\lambda_L\, (e^+\bar d)$ & 1 \\[-4ex]
       & $S_{3L}$        
       & {\bf 45} & $\left\{\rule{0pt}{30pt}\right.$
       & $1/3$& $-\lambda_L\, (e^+\bar u)$,
$-\lambda_L\, (\bar\nu\bar d)$ & $1/2$ \\[-4ex]
       &         &        
       &         & $-2/3$& $\sqrt 2\lambda_L\, (\bar\nu\bar u)$ & 0 \\[2ex]
\raisebox{-2ex}[0pt]{$F=0$} &  
\raisebox{-2ex}[0pt]{$R_{2L}$} &        
\raisebox{-2ex}[0pt]{{\bf 45}} &
\raisebox{-2ex}[0pt]{$\left\{\rule{0pt}{20pt}\right.$} 
& $5/3$ & $\lambda_L\, (e^+u)$ & 1 \\[-4ex]
       &  &  &        & $2/3$ & $\lambda_L\, (\bar\nu u)$ & 0 \\[2ex]
       & 
\raisebox{-2ex}[0pt]{$R_{2R}$} &
\raisebox{-2ex}[0pt]{{\bf 45}} & 
\raisebox{-2ex}[0pt]{$\left\{\rule{0pt}{20pt}\right.$}
& $5/3$ & $\lambda_R\, (e^+u)$ & 1 \\[-4ex]
  & &    &  & $2/3$ & $-\lambda_R\, (e^+d)$ &  1\\[2ex]
       & 
\raisebox{-2ex}[0pt]{$\widetilde R_{2L}$} & 
\raisebox{-2ex}[0pt]{{\bf 10/15}} & 
\raisebox{-2ex}[0pt]{$\left\{\rule{0pt}{20pt}\right.$} &
$2/3$ &$\lambda_L\, (e^+d)$ & 1 \\[-4ex] 
       &       &    &         & $-1/3$& $\lambda_L\, (\bar\nu d)$ 
&  0 \\[1.2ex] \hline
Vectors & & & & & \\ \hline
\rule{0pt}{15pt}
\raisebox{-3ex}[0pt]{$F=-2$} & 
\raisebox{-2ex}[0pt]{$V_{2L}$} & 
\raisebox{-2ex}[0pt]{{\bf 24}} & 
\raisebox{-2ex}[0pt]{$\left\{\rule{0pt}{20pt}\right.$} &
 $4/3$ & $\lambda_L\, (e^+\bar d)$ & 1 \\[-4ex]
       & & && $1/3$ & $\lambda_L\, (\bar\nu\bar d)$ & 0 \\[2ex]
       & 
\raisebox{-2ex}[0pt]{$V_{2R}$} & 
\raisebox{-2ex}[0pt]{{\bf 24}} &
\raisebox{-2ex}[0pt]{$\left\{\rule{0pt}{20pt}\right.$} & 
$4/3$ & $\lambda_R\, (e^+\bar d)$ & 1 \\[-4ex]
       &   &              &         &
        $1/3$ & $\lambda_R\, (e^+\bar u)$ & 1 \\[2ex]
       & 
\raisebox{-2ex}[0pt]{$\widetilde V_{2L}$} & 
\raisebox{-2ex}[0pt]{{\bf 10/15}} &
\raisebox{-2ex}[0pt]{$\left\{\rule{0pt}{20pt}\right.$} &
 $1/3$ & $\lambda_L\, (e^+\bar u)$ & 1 \\[-4ex]
       &       &          &         &
        $-2/3$ & $\lambda_L\, (\bar\nu\bar u)$ & 0 \\[2ex]
$F=0$  & $U_{1L}$  & {\bf 10} && $2/3$ & $\lambda_L\, (e^+d)$,
$\lambda_L\, (\bar\nu u)$ & $1/2$ \\
       & 
$U_{1R}$  & {\bf 10} &&  $2/3$ & $\lambda_R\, (e^+\bar d)$ & 1 \\
       & 
$\widetilde U_{1R}$ & 
{\bf 75} &&
 $5/3$ & $\lambda_R\, (e^+u)$ & 1 \\[2ex]
       &    &             &         &
        $5/3$ & $\sqrt 2\lambda_L\, (e^+u)$ & 1 \\[-4ex]
       & 
$U_{3L}$  & 
{\bf 40} &
$\left\{\rule{0pt}{30pt}\right.$ &
 $2/3$ & $-\lambda_L\, (e^+d)$, $\lambda_L\, (\bar\nu u)$ & $1/2$\\[-4ex]
       &      &           &         &
        $-1/3$ & $\sqrt 2\lambda_L\, (\bar\nu d)$ & 0 \\[2ex]
\hline\hline
\end{tabular}
\end{center}
\label{lqprop}
\end{table}}

Attempts to explain the HERA data have led to new types of LQs not obtainable 
from the BRW analysis; this occurs as follows. 
Since the HERA excess appears in the $e^+p$ channel and not in $e^-p$ we 
would be forced to conclude that our HERA associated LQ has $F=0$. Furthermore, 
since vector LQs have a far larger cross section at the Tevatron than do 
scalars{\cite {old}} 
and, since neither CDF nor D0 have observed LQ pair 
production{\cite {cdfd0}}, we must conclude the HERA LQ is a scalar. 
One of the difficulties associated with the LQ interpretation of the HERA 
excess is then immediately obvious from Table~\ref{lqprop} in that 
all $F=0$ scalars 
have a branching fraction($B_\ell$) into $ej$ of unity, something already 
excluded by CDF/D0 searches if the LQ mass is anywhere near 200 GeV. This 
necessitates the construction of LQ models that go beyond{\cite {us}} those 
considered by BRW and this can only be done by dropping (at least) one of the 
BRW assumptions, \eg, assumption (f). 
Thus we should remember that the BRW list is not necessarily an 
exhaustive one and be prepared for LQs with other possible quantum numbers. 
The BRW set does, however, provide a fertile testing ground for the ability 
of colliders to distinguish the various possibilities from one another.

\section{Leptoquark Pair Production}

As is well known, though hadron colliders provide a high mass reach for 
searches they are incapable of determining the electroweak quantum numbers of 
the LQs once they are discovered. However, it is possible that from the 
cross section and angular 
distributions the spin of the LQ may be determined. Before the turn on of a 
first generation lepton collider it is likely that the LHC will have probed 
the LQ mass range up to $\simeq 1.3-1.5$ TeV for scalars and $2.1-2.5$ TeV for 
vectors{\cite {me}} in the pair production channel. 
(Here we have accounted for the new NLO pair production cross 
section results obtained by Kr\"amer \etal {\cite {kramer}}). Thus we should 
already know if a LQ will be kinematically accessible at the first generation 
lepton collider. 

\begin{figure}[htbp]
\centerline{
\epsfig{figure=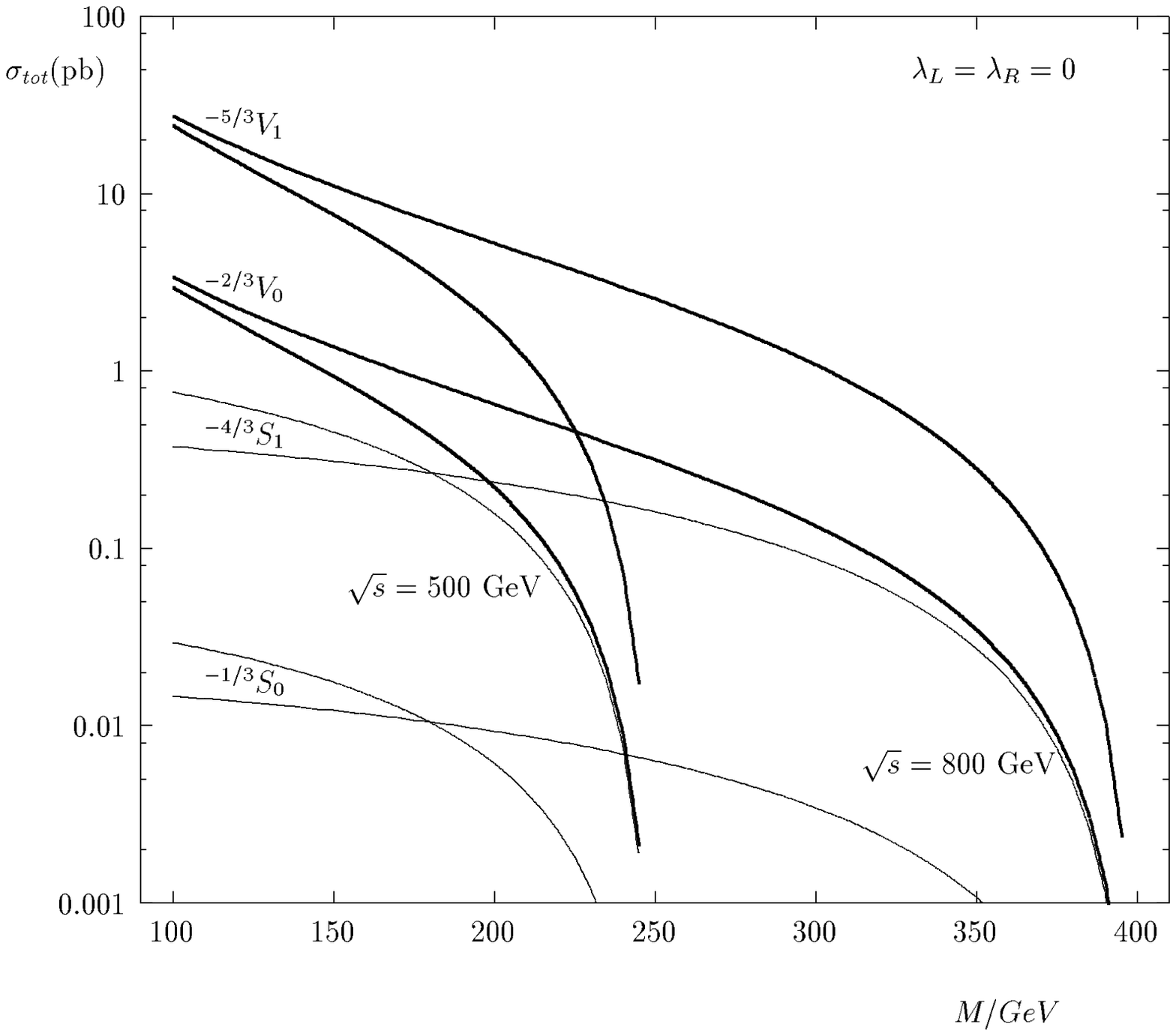,height=15.5cm,width=14cm,angle=0}}
\vspace*{-4.5cm}
\fcaption{ Total cross sections for leptoquark pair-production at
  fixed center-of-mass energies as a function of the leptoquark mass
  $M$ assuming vanishing Yukawa couplings and including corrections
  due to beamstrahlung and ISR from the analysis of RSS. Here S or V 
  refer to the LQ spin, the upper(lower) index is the corresponding 
  electric charge(weak isospin).}
\label{figsigtot}
\end{figure}

At planned lepton colliders, which have smaller values of $\sqrt s$ than the 
LHC, the pair production cross section and angular distribution 
in $e^+e^-$ or $\mu^+\mu^-$ collisions alone already tells us much about 
both the LQ 
spin and electroweak quantum numbers. Fig.~\ref{figsigtot} from the recent 
analysis of R\"uckl, Settles and Spiesberger(RSS){\cite {rss}}, shows that LQ 
pair production cross sections at lepton colliders 
are large and give quite reasonable rates 
assuming canonical luminosities in the 50-100 $fb^{-1}$ range. 
(We note that these results assume that the strength of the Yukawa 
coupling, $\lambda$, at the 
$eq$ vertex is quite weak in comparison to electroweak couplings. If this is 
{\it not} the case then LQ pair production will also occur through 
$t(u)-$channel 
quark exchange as well as $s-$channel $\gamma$ and $Z$ exchange. This can 
result in a significant change in the overall production rate as well as the 
angular distribution.) We also see 
from this figure the rather strong variations in the cross section due to both 
LQ spin and quantum number choices.

This sensitivity can be seen more clearly from Table~\ref{lqnlc} which shows 
the cross sections and polarization asymmetries for all the scalar LQs in 
the BRW scheme given in  
Table\ref{lqprop} assuming degenerate multiplets. It is easy to see that from 
the cross section and polarization asymmetry it will be quite simple to 
distinguish the various models. We remind the reader again that there can be 
more LQ quantum number assignments than are obtainable from simply following 
the BRW assumptions{\cite {us}} and some confusion may thus occur in quantum 
number extractions if care is not exercised.

\vspace*{-0.5cm}
\nn
\vspace*{0.4mm}
\begin{table*}[htbp]
\tcaption{ Cross sections for the three scalar 
leptoquark pair decay channels in fb at 
a 500 GeV NLC assuming complete leptoquark multiplets with a common mass of 
200 GeV. The polarization asymmetry in the $\ell \ell jj$ channel is also 
given. In all cases $\tilde \lambda \ll 1$ is assumed.}
\leavevmode
\begin{center}
\label{lqnlc}
\begin{tabular}{lcccc}
\hline
\hline
Leptoquark&$\ell \ell jj$&$\ell\nu jj$&$\nu\nu jj$&$A^{LR}_{\ell \ell jj}$ \\
\hline

$S_{1L}$        & 1.88 & 3.77 & 1.88 & -0.618 \\
$S_{1R}$        & 7.53 & 0.0  & 0.0  & -0.618 \\
$\widetilde S_{1R}$ &120.4 & 0.0  & 0.0  & -0.618 \\
$S_{3L}$        &192.2 & 3.77 & 1.88 &  0.931 \\
$R_{2L}$        &181.0 & 0.0  & 80.4 &  0.196 \\
$R_{2R}$        &261.4 & 0.0  & 0.0  & -0.141 \\
$\widetilde R_{2L}$ & 47.6 & 0.0  & 33.2 &  0.946 \\
\hline
\hline
\end{tabular}
\end{center}
\end{table*}

The search reach for LQs in the pair production mode in $e^+e^-$ collisions 
has been most thoroughly 
studied in the recent RSS analysis. These authors perform a very detailed 
simulation study including all SM backgrounds, detector cuts and 
smearing, decay and hadronization effects as well as including beamstrahlung 
and initial state radiation(ISR). Table~\ref{tablimits} from the RSS analysis 
shows the expected search reach for all BRW type LQs. In almost all cases we 
see that the reach approaches the kinematic limit for pair production of 
$\simeq \sqrt s/2$ in both the $eejj$ and $e\nu jj$ mode. The reach is seen 
to be usually somewhat less in the $\nu\nu jj$ channel. It appears that LQs 
will not be missed at a lepton collider. 

\begin{table}[hbtp]
\tcaption{\it Discovery limits for leptoquarks (masses in GeV) at 
  $\protect\sqrt{s} = 500$ GeV (${\cal L} = 20$ fb$^{-1}$) and
  $\protect\sqrt{s} = 800$ GeV (${\cal L} = 50$ fb$^{-1}$) requiring a
  5$\sigma$ effect from the analysis of RSS. I, II and III label the channels 
  $eejj$, $e\nu jj$ and $\nu \nu jj$ respectively. $N_{bg}$ is the number of 
  background events passing the cuts in a given channel. 
  Dashes indicate cases where the corresponding search
  is not possible, $*$ means no sensitivity to masses above 100 GeV
  with the cuts considered.}
\begin{center}
\begin{tabular}{|l|l|c|c|c|c|c|c|}
\hline  \multicolumn{2}{|l|}{\rule{0mm}{5mm}}
      & \multicolumn{3}{|c|}{$\sqrt{s}=500$ GeV}
      & \multicolumn{3}{|c|}{$\sqrt{s}=800$ GeV} \\[1mm]
\hline  \multicolumn{2}{|r|}{\rule{0mm}{4.5mm} Search}      
      &   I &    II  &  III   &   I &    II  &   III \\
\hline
\multicolumn{2}{|r|}{\rule{0mm}{5mm} $5\sqrt{N_{bg}}$}
      &  18 &    61  &  251   &  21 &    60  &   375 \\[1mm]
\hline
States & $B_{eq}$ &
\multicolumn{6}{|c|}{\rule{0mm}{5mm} Mass reach in GeV} \\[1mm]
\hline
\hline \rule{0mm}{5mm}
${}^{-1/3}S_0$ & $2/3$ 
      &\hspace{1mm}202\hspace{1mm}
            &   $*$  &   $*$ &\hspace{1mm}318\hspace{1mm}
                                    &   $*$  & $*$ 
\\
               & $1/2$ 
      & 183 &   $*$  &   $*$  & 289 &   $*$  & $*$ 
\\
              & $1$ 
      & 217 &   --   &    --  & 350 &    --  &  -- 
\\[1mm]
\hline \rule{0mm}{5mm}
${}^{-4/3}\tilde{S}_0$ & $1$ 
      & 242 &   --   &    --  & 387 &    --  &  -- 
\\[1mm]
\hline \rule{0mm}{5mm}
${}^{2/3}S_1$ & $0$ 
      & -- &    --   &   225  &  -- &    --  &  275
\\[1mm]
\hline \rule{0mm}{5mm}
${}^{-1/3}S_1$ & $1/2$ 
      & 183 &   $*$  &   $*$  & 289 &    $*$ & $*$ 
\\[1mm]
\hline \rule{0mm}{5mm}
${}^{-4/3}S_1$ & $1$ 
      & 244 &   --   &    --  & 389 &    --  &  --
\\[1mm]
\hline \rule{0mm}{5mm}
${}^{-2/3}S_{1/2}$ & $1/2$ 
      & 230 &  221   &   179  & 369 &   359  &  $*$
\\
                   & $0$ 
      & --  &   --   &   218  &  -- &    --  &  239
\\
                   & $1$ 
      & 240 &   --   &   --   & 384 &    --  &  --
\\[1mm]
\hline \rule{0mm}{5mm}
${}^{-5/3}S_{1/2}$ & $1$
      & 244 &  --    &   --   & 389 &   --   &  --  
\\[1mm]
\hline \rule{0mm}{5mm}
${}^{1/3}\tilde{S}_{1/2}$ & $0$
      &  -- &  --    &  198   & --  &   --   & 146 
\\[1mm]
\hline \rule{0mm}{5mm}
${}^{-2/3}\tilde{S}_{1/2}$ & $1$
      & 237 &  --    &   --   & 379 &   --   &  --  
\\[1mm]
\hline
\hline \rule{0mm}{5mm}
${}^{-1/3}V_{1/2}$ & $1/2$ 
      & 241 &  237   &  220   & 385 &   380  & 266 
\\
                   & $0$ 
      &  -- &   --   &  236   &  -- &    --  & 326
\\
                   & $1$ 
      & 245 &   --   &   --   & 392 &    --  &  -- 
\\[1mm]
\hline \rule{0mm}{5mm}
${}^{-4/3}V_{1/2}$ & $1$
      & 247 &   --   &   --   & 395 &    --  &  -- 
\\[1mm]
\hline \rule{0mm}{5mm}
${}^{2/3}\tilde{V}_{1/2}$ & $0$
      & --  &   --   &  236   & --  &    --  & 326
\\[1mm]
\hline \rule{0mm}{5mm}
${}^{-1/3}\tilde{V}_{1/2}$ & $1$
      & 244 &   --   &   --   & 390 &    --  &  -- 
\\[1mm]
\hline \rule{0mm}{5mm}
${}^{-2/3}V_0$ & $2/3$
      & 241 &  233   &  195   & 385 &   373  & 200
\\
               & $1/2$
      & 238 &  234   &  212   & 380 &   376  & 244
\\
               & $1$
      & 244 &   --   &   --   & 390 &    --  &  -- 
\\[1mm]
\hline \rule{0mm}{5mm}
${}^{-5/3}\tilde{V}_0$ & $1$
      & 247 &   --   &   --   & 396 &    --  &  --  
\\[1mm]
\hline \rule{0mm}{5mm}
${}^{1/3}V_1$ & $0$
      & --  &   --   &  241   & --  &    --  & 352
\\[1mm]
\hline \rule{0mm}{5mm}
${}^{-2/3}V_1$ & $1/2$
      & 238 &  234   &  212   & 380 &   375  & 244
\\[1mm]
\hline \rule{0mm}{5mm}
${}^{-5/3}V_1$ & $1$
      & 248 &   --   &   --   & 396 &    --  &  --  
\\[1mm]
\hline
\end{tabular} 
\label{tablimits}
\end{center}
\end{table}

%
\vspace*{-0.5cm}
\nn
\begin{figure}[htbp]
\centerline{
\epsfig{figure=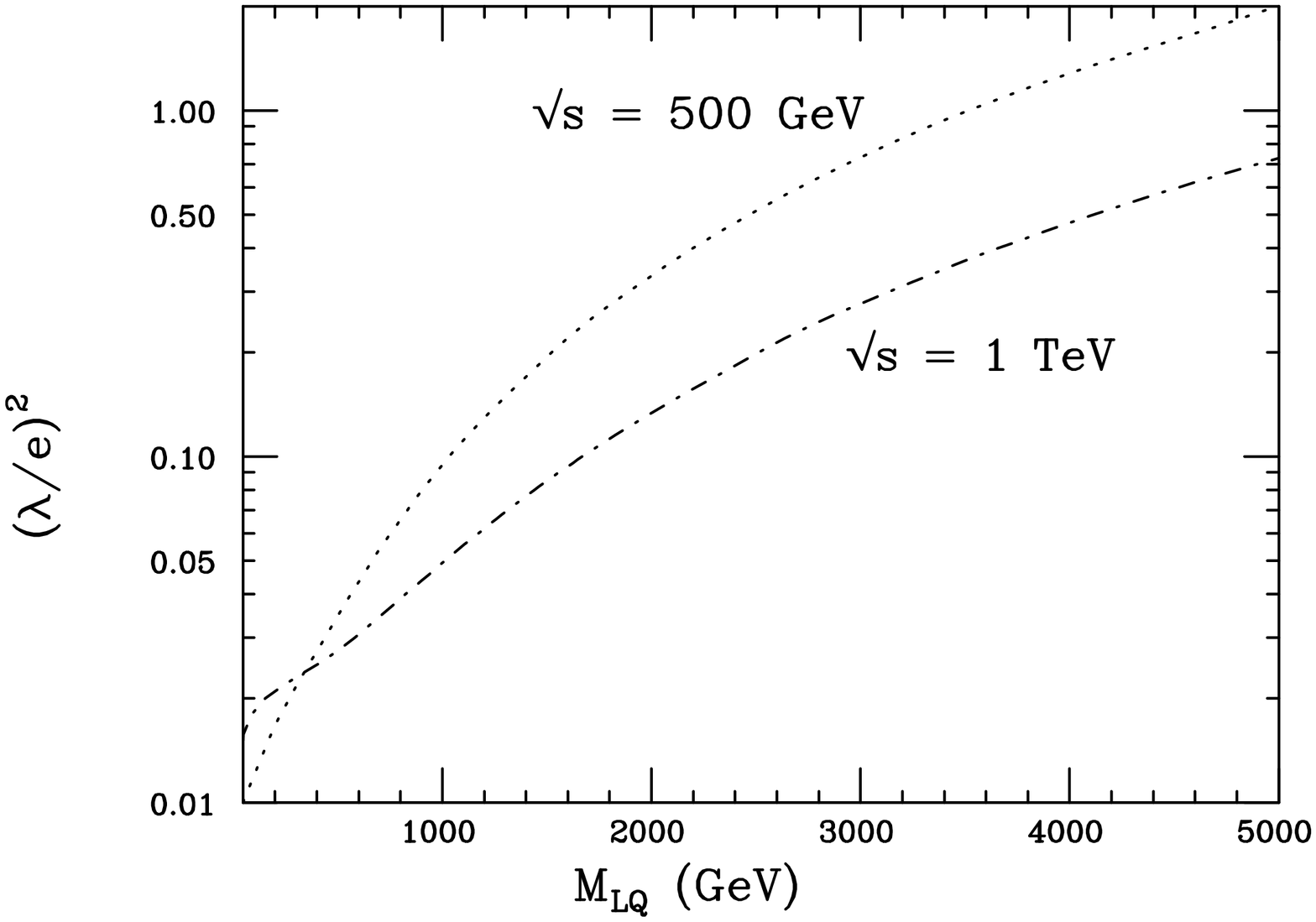,height=10cm,width=12cm,angle=-90}}
\vspace*{-0.5cm}
\fcaption{$95\%$ CL indirect discovery region(to the left of the curves) 
at a 500 GeV and 1 TeV $e^+e^-$ collider with integrated luminosities of 
50 and 100 $fb^{-1}$ from the analysis of Hewett.}
\label{joafig}
\end{figure}
\vspace*{0.4mm}

It may also be possible to {\it indirectly} probe the existence of LQs in 
$e^+e^- \to q\bar q$ processes since $t$-channel LQ exchange can contribute 
significantly--provided the Yukawa coupling is sufficiently large. Here one 
looks for deviations in both the total cross section as well as the angular 
distribution in a manner resulting from the $t$-channel exchange. 
Fig.~\ref{joafig} from the updated analysis of Hewett{\cite {early}}, a part 
of the DPF study in Ref.~3, shows the region in LQ mass vs LQ Yukawa 
coupling space that can be probed by this technique for the case of the 
$E_6$-type LQs, \ie, $S_{1L,R}$. Note that even for 
$(\lambda/e)^2=0.1$ the reach is of order $2\sqrt s$ which is more than 
4 times that obtainable from direct pair production. Comparable reaches are 
also obtainable for other LQ types. 
The OPAL Collaboration{\cite {opal}} has recently performed this type of 
analysis with real data at LEPII and have obtained interesting constraints.

\section{Single Leptoquark Production}

\begin{figure}[htbp]
\centerline{
\psfig{figure=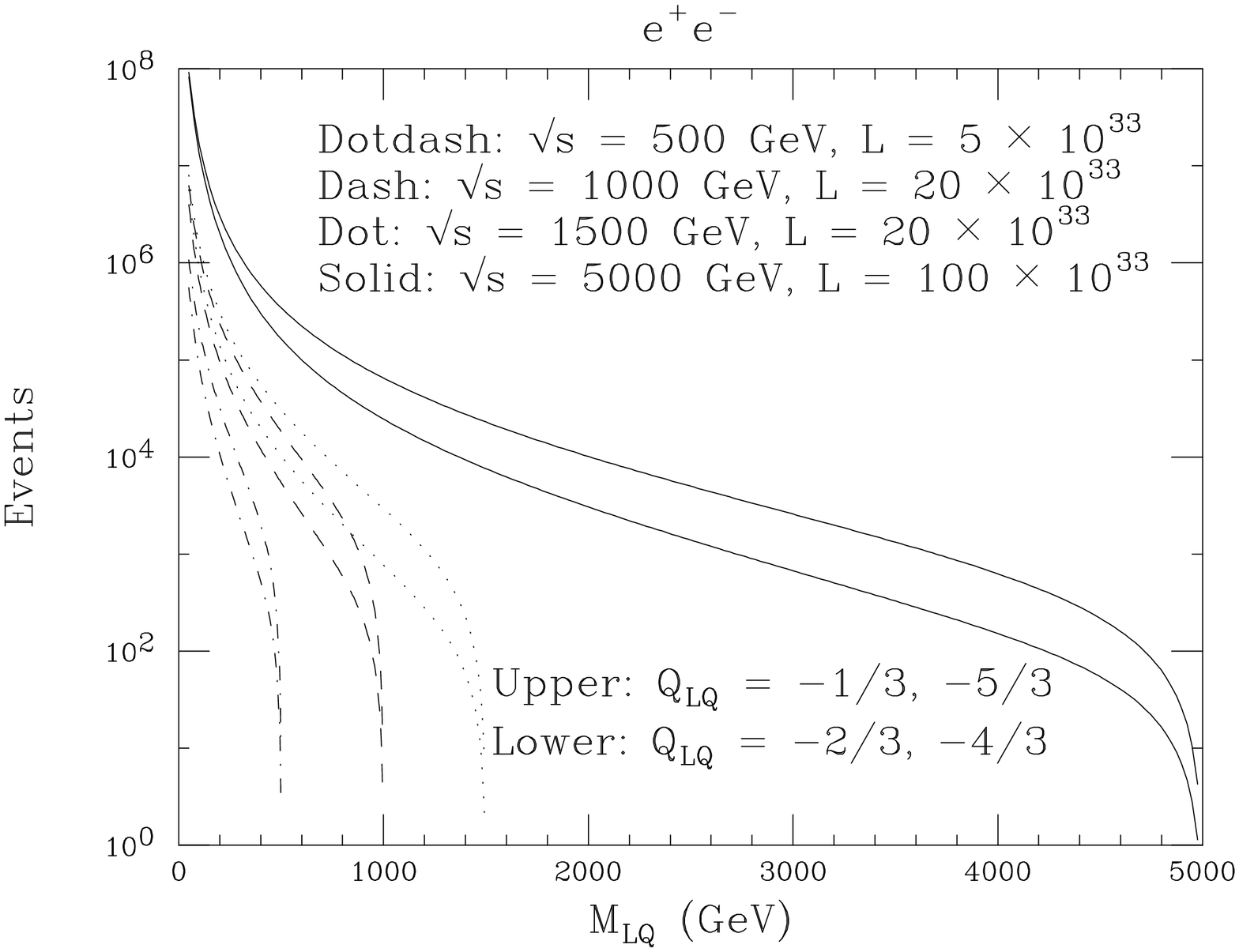,height=6.0cm,width=6.0cm,angle=90}
\hspace*{0mm}
\psfig{figure=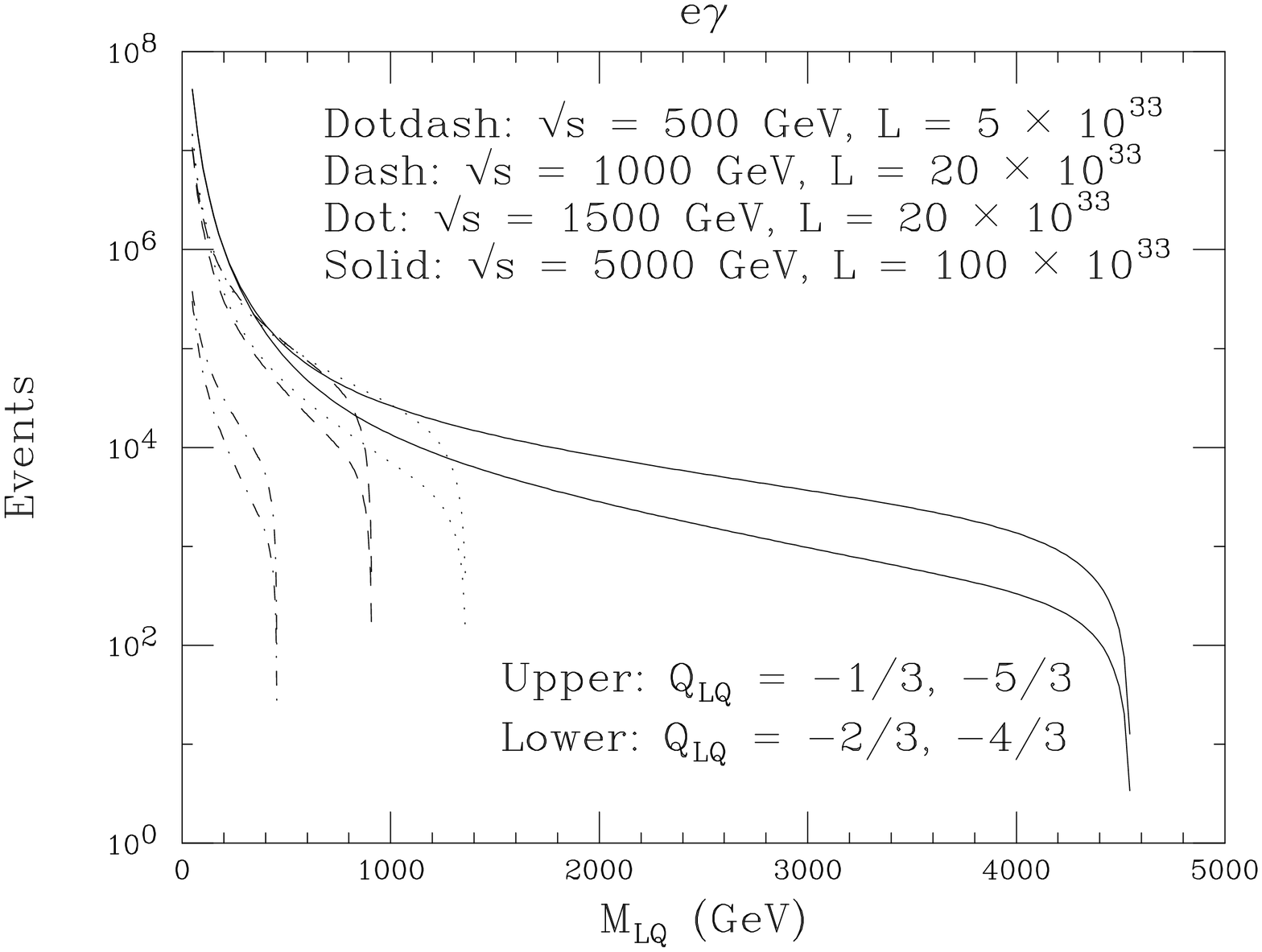,height=6.0cm,width=6.0cm,angle=90}}
\vspace*{0.2cm}
\centerline{
\psfig{figure=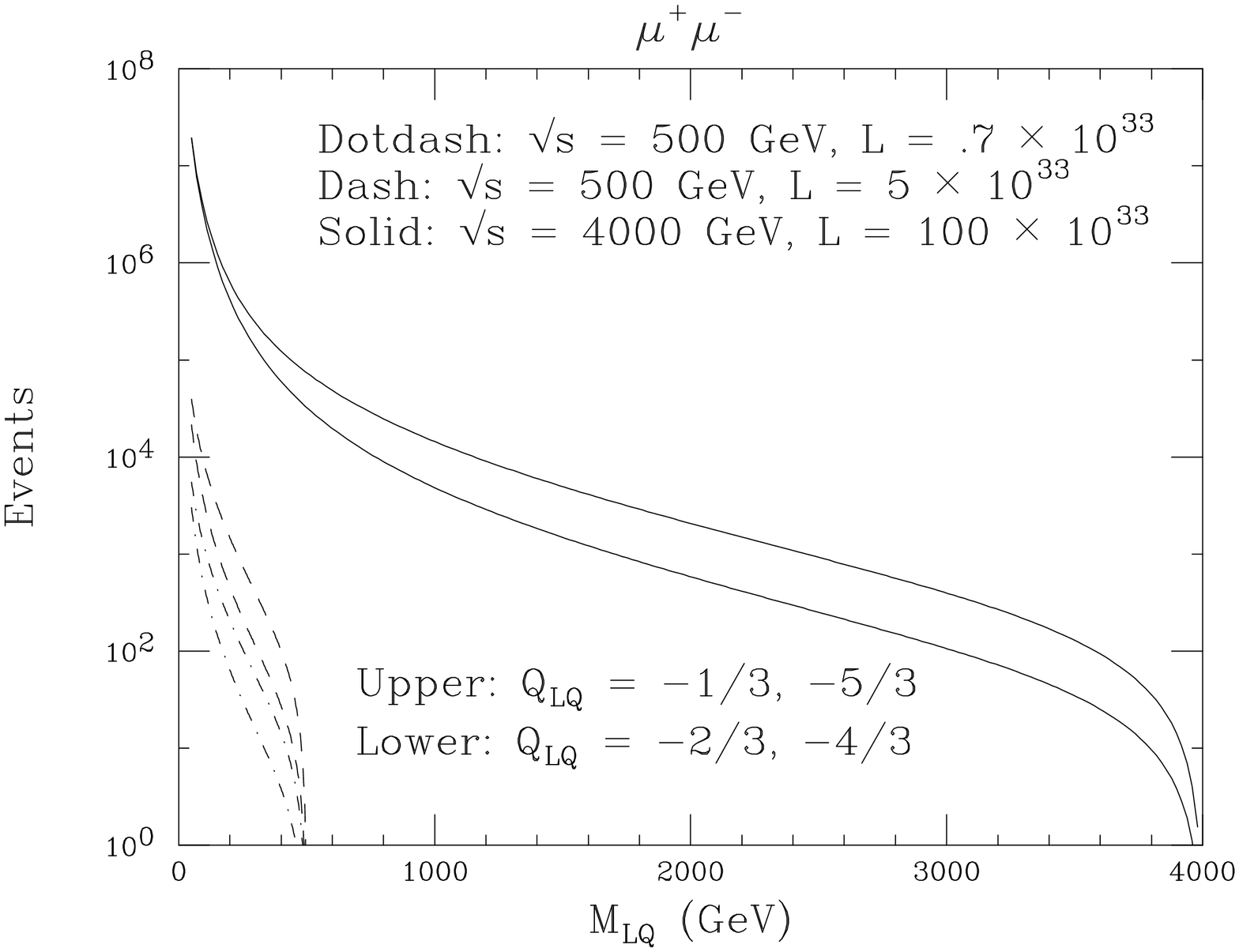,height=6.0cm,width=6.0cm,angle=90}}
\vspace*{0.2cm}
\fcaption{Event rates for single leptoquark production in
$e^+e^-$(WW), $e\gamma$(BL), and $\mu^+\mu^-$(WW) collisions from Doncheski and 
Godfrey. The center of mass energies and integrated luminosities are given 
by the line labelling in the figures.}
\label{steve1}
\end{figure}

In order to directly probe beyond the kinematic reach associated with pair 
production it is necessary to consider single LQ production which occurs via 
$\gamma \ell$ collisions{\cite {egam}} and whose rate depends on the square 
of the Yukawa coupling, $\lambda$. The photon in this case can either be the 
result of Weizs\"acker-Williams(WW) emission at a conventional lepton collider 
or it arises from a backscattered laser(BL) beam and both possibilities have 
been examined in the literature. One of the most complete studies of both 
these processes has recently been performed by Doncheski and 
Godfrey{\cite {egam,steve}}. Fig.~\ref{steve1} from that analysis 
shows the event rate for these processes assuming electromagnetic strength 
Yukawa couplings for a number of different collider options. Table 4 
shows the corresponding search reach in each of these cases obtained by the  
same authors under the same set of assumptions. An important ingredient in 
this analysis is the inclusion of contributions to the cross section 
which arise due to the hadronic content of the photon; specifically, 
Doncheski and Godfrey make use of the parton 
densities of Gl\"uck, Reya and Vogt(GRV){\cite {GRV}}. Note that the 
contribution from these subprocesses are uncertain by about a factor of two 
given the uncertainties in the photon distribution functions. 

\vspace*{-0.5cm}
\nn
%

\begin{table*}[htpb]
\begin{center}
\tcaption{Leptoquark discovery limits for $e^+e^-$(WW), 
$e\gamma$(BL), and $\mu^+\mu^-$(WW) colliders from the analysis of Doncheski 
and Godfrey.  The discovery limits are based 
on the production of 100 LQ's for the centre of mass energies 
and integrated luminosities given in columns one and two.
The results were obtained using the GRV distribution functions and the LQs 
are labeled by their electric charges.}
\begin{tabular}{llllll}
\hline
\hline
\multicolumn{6}{c}{$e^+e^-$ Colliders}\\ \hline
$\sqrt{s}$ (TeV) & $L$ (fb$^{-1}$) & \multicolumn{2}{c}{Scalar} & 
\multicolumn{2}{c}{Vector}  \\ \hline 
 & & -1/3, -5/3 & -4/3, -2/3 & -1/3, -5/3 & -4/3, -2/3 \\ \hline 
 0.5  & 50 & 490 & 470 & 490 & 480 \\
 1.0  & 200 & 980 & 940 & 980 & 970 \\
 1.5 & 200 & 1440 & 1340 & 1470 & 1410  \\
 5.0 & 1000 & 4700 & 4200 & 4800 & 4500  \\ \hline \hline
\end{tabular}
\vskip 0.6cm
\begin{tabular}{llllll}
\hline
\hline
\multicolumn{6}{c}{$e\gamma$ Colliders} \\ \hline
$\sqrt{s}$ (TeV) & $L$ (fb$^{-1}$) & \multicolumn{2}{c}{Scalar} & 
\multicolumn{2}{c}{Vector}  \\ \hline 
 & & -1/3, -5/3 & -4/3, -2/3 & -1/3, -5/3 & -4/3, -2/3 \\ \hline 
 0.5  & 50 & 450 & 450 & 450 & 440 \\
 1.0  & 200 & 900 & 900 & 910 & 910 \\
 1.5 & 200 & 1360 & 1360 & 1360 & 1360  \\
 5.0 & 1000 & 4500 & 4400 & 4500 & 4500  \\ \hline \hline
\end{tabular}
\vskip 0.6cm
\begin{tabular}{llllll}
\hline
\hline
\multicolumn{6}{c}{$\mu^+\mu^-$ Colliders} \\ \hline
$\sqrt{s}$ (TeV) & $L$ (fb$^{-1}$) & \multicolumn{2}{c}{Scalar} & 
\multicolumn{2}{c}{Vector} \\ \hline 
 & & -1/3, -5/3 & -4/3, -2/3 & -1/3, -5/3 & -4/3, -2/3 \\ \hline 
 0.5  & 0.7 & 250 & 170 & 310 & 220 \\
 0.5  & 50 & 400 & 310 & 440 & 360 \\
 4.0 & 1000 & 3600 & 3000 & 3700 & 3400  \\ \hline
\hline 
\end{tabular}
\end{center}
\label{steve2}
\end{table*}
\begin{table*}[htbp]
\tcaption{Rates for single scalar 
leptoquark production in $\gamma e$ collisions at 
a 500 GeV NLC assuming complete leptoquark multiplets with a common mass of 
200 GeV and an integrated luminosity of $50fb^{-1}$. In all cases 
$\lambda/e =0.1$ is assumed and a $p_T$ cut on the quark jet of 10 GeV 
has been applied. The charged lepton branching fraction for the produced 
multiplet is also given.}
\leavevmode
\begin{center}
\label{lqs}
\begin{tabular}{lccc}
\hline
\hline
Leptoquark& Backscattered Laser & Weizs\"acker-Williams & $B_{\ell}$ \\
\hline

$S_{1L}$        &      212.  &  56.8  & 0.5 \\
$S_{1R}$        &      212.  &  56.8  & 1   \\
$\widetilde S_{1R}$ &  109.  &  24.4  & 1   \\
$S_{3L}$        &      430.  &  106.  &$\simeq 0.75$ \\
$R_{2L}$        &      332.  &  79.5  & 1\\
$R_{2R}$        &      381.  &  92.6  & 1\\
$\widetilde R_{2L}$ &  49.1  &  13.1  & 1\\
\hline
\hline
\end{tabular}
\end{center}
\end{table*}

A similar analysis in Table~\ref{lqs} from Ref.~6 directly 
compares the production rates obtainable via the WW and BL processes. 
From either of these 
analyses we see that the search reach in the $\gamma \ell$ mode is essentially 
$\simeq 0.9\sqrt s$ for Yukawas of strength $0.1e$ or greater, independently 
of the LQ type.

Is it possible to distinguish the various LQ types in the $\gamma e$ mode once 
they are observed? Doncheski and Godfrey{\cite {egam,steve}} have performed 
a very extensive analysis in the attempt to answer this question. Here, since 
the Yukawa coupling itself is {\it a priori} unknown, the 
cross section measurement itself cannot be used to identify the LQ, and we are 
thus forced to rely only upon various asymmetries to probe LQ properties. The 
first asymmetry examined by Doncheski and Godfrey employs only the electron 
beam polarization and corresponds to the conventional left-right asymmetry 
given by: 
\begin{displaymath}
A^{+-} = {{\sigma^+ - \sigma^-}\over{\sigma^+ + \sigma^-}}
= {{\lambda_L^2 -\lambda_R^2}\over {\lambda_L^2 + \lambda_R^2}}.
\end{displaymath}
Assuming that LQs are chirally coupled this divides the various models in the 
BRW classification
into three distinct bins depending on whether the particular LQ type couples 
only to the left- or right-handed electron or can couple to either helicity: 
\begin{description}
\item[$e^-_L$:] $\tilde{R}_{2L}$, $S_{3L}$, $U_{3L}$, $\tilde{V}_{2L}$
\item[$e^-_R$:] $\tilde{S}_{1R}$, $\tilde{U}_{1R}$
\item[$e^-_U$:] $U_{1L,R}$, $V_{2L,R}$, $R_{2L,R}$, $S_{1L,R}$
\end{description}
Here ``$e^-_U$'' means this LQ type can couple to either helicity as shown in 
Table~\ref{lqprop}. Of course, this is just convention since the 
polarization will pick out just one of these two states.

Doncheski and Godfrey showed that one can go further and trivially distinguish 
whether the LQs are scalar or 
vector. This could be accomplished in two ways.  In the first case one can 
study the angular distributions of the leptoquark decay products.  In the 
second they employ the double polarization asymmetry: 
\begin{displaymath}
A_{LL} = {{ (\sigma^{++} + \sigma^{--} ) - (\sigma^{+-} + 
\sigma^{-+} )} \over
{ (\sigma^{++} + \sigma^{--} ) + (\sigma^{+-} + \sigma^{-+} )}}
\end{displaymath}
where the first index refers to the electron helicity and the second 
to the quark helicity, the quark arising from the internal structure of the 
photon.  Because scalars only have a non-zero cross 
section for $\sigma^{++}$ and $\sigma^{--}$, for scalar LQs one finds that 
the parton level asymmetry for $e q$ collisions is $\hat{a}_{LL}= +1$.  
Similarly, since vectors only have a 
non-zero cross section for $\sigma^{+-}$ and $\sigma^{-+}$ for vector 
LQs $\hat{a}_{LL}=-1$.  To obtain the observable
asymmetries we convolute the parton level cross sections with the 
polarized photon distribution functions.  Doing so will reduce the asymmetries 
from their parton level values of $\pm 1$ so one 
must determine whether the observable asymmetries resulting from this 
convolution can be used to distinguish 
between the leptoquark types.  The complete expressions for the double 
longitudinal spin asymmetry $A_{LL}$ are given in Doncheski and 
Godfrey{\cite {egam}}. These authors show in detail that out to masses of 
order $\simeq 0.75\sqrt s$, vector and scalar LQs are easily distinguished as 
are LQs with the same chirality of their couplings.

\section{Leptoquarks in \boldmath$e^-e^-$ Collisions}

LQs might also be pair produced in $e^-e^-$ collisions provided one's model is 
sufficiently complex{\cite {fram}}. (It is obvious that an $e^-e^-$ 
collider can also be run in the $\gamma e$ mode with the advantage that both 
beams can be polarized.) Since the initial state in $e^-e^-$ collisions has 
$L=2$, the two LQs produced in the final state must be distinct 
in that one must have $F=0$ and the other will have $F=2$. This implies that 
any model that predicts LQ production in this channel must have LQs of more 
than one species, such as the $SU(15)$ or the 331 models of 
Frampton{\cite {fram}}, with the production resulting from $t-$ or 
$u-$channel quark exchange. In some models, such as the 331 case, an 
$s-$channel exchange of a $L=2$ gauge boson also contributes; depending upon 
mass relations this exchange may be resonant and lead to a very large cross 
section. If the two LQs 
have different masses, $M_1<M_2$, this mode may allow an extension of the 
search reach obtainable in $e^+e^-$ collisions provided 
$M_1+M_2< \sqrt s <2M_2$.

\section{Summary and Outlook}

Lepton colliders in the $e^+e^-/\mu^+\mu^-$, $\gamma e$ and $e^-e^-$ modes 
provide unique ways to both discover LQs of all types {\it and} to obtain 
detailed information about their electroweak quantum numbers--something not 
possible at present or future hadron colliders. 
The phenomenology of LQ models is particularly rich, particularly if one 
goes beyond the BRW scheme. Analyses have become increasingly complex and have 
evolved in sophistication to the point where detector considerations are 
becoming increasingly important. Although much work has been done, 
there is still a lot of work to be done in the future.

\def\MPL #1 #2 #3 {Mod. Phys. Lett. {\bf#1},\ #2 (#3)}
\def\NPB #1 #2 #3 {Nucl. Phys. {\bf#1},\ #2 (#3)}
\def\PLB #1 #2 #3 {Phys. Lett. {\bf#1},\ #2 (#3)}
\def\PR #1 #2 #3 {Phys. Rep. {\bf#1},\ #2 (#3)}
\def\PRD #1 #2 #3 {Phys. Rev. {\bf#1},\ #2 (#3)}
\def\PRL #1 #2 #3 {Phys. Rev. Lett. {\bf#1},\ #2 (#3)}
\def\RMP #1 #2 #3 {Rev. Mod. Phys. {\bf#1},\ #2 (#3)}
\def\ZPC #1 #2 #3 {Z. Phys. {\bf#1},\ #2 (#3)}
\def\IJMP #1 #2 #3 {Int. J. Mod. Phys. {\bf#1},\ #2 (#3)}
\nonumsection{References}


\begin{thebibliography}{99}
\bibitem{hera}
C. Adloff \etal, H1 Collaboration, \ZPC C74 191 1997 .
J. Breitweg \etal, ZEUS Collaboration, \ZPC C74 207 1997 .
%
\bibitem{brw}
W. Buchm\" uller, R. R\" uckl, and D. Wyler, \PLB B191 442 1987 ;
S. Davidson, D. Bailey, and B.A. Campbell, \ZPC C61 613 1994 ;
M. Leurer, \PRD D50 536 1994 , and \PRD D49 333 1994 . 
%
\bibitem{rev}
For a complete set of references and a review of the physics of 
leptoquarks at colliders see, 
A. Djouadi, J. Ng and T.G. Rizzo, SLAC-PUB-95-6772, 1995, 
a part of the DPF long-range planning study, published in
{\it Electroweak Symmetry Breaking and Physics Beyond the Standard Model}, 
eds. T.\ Barklow, S.\ Dawson, H.\ Haber, and J.\ Seigrist (World Scientific 
1996). See also, J.L. Hewett and T.G. Rizzo, \PR  183 193  1989 .
%
\bibitem {old}
J.L. Hewett, T.G. Rizzo, S. Pakvasa, H.E. Haber, and A. Pomarol, in
{\it Proceedings of the Workshop on Physics at Current Accelerators and
the Supercollider}, June 1993, ed. J.L. Hewett, A. White, and D. Zeppenfeld.
%
\bibitem{cdfd0}
F. Abe, CDF Collaboration, hep-ex/9708017, submitted to Phys. Rev. Lett. 
See also, H.S Kambara, CDF Collaboration, talk given at the {\it $12^{th}$ 
Workshop on Hadron Collider Physics}, Stony Brook, NY June 5-11, 1997. 
B. Abbott, D0 Collaboration, hep-ex/9707033, submitted to Phys. Rev. Lett. See 
also, D. Norman, D0 Collaboration, talk given at the {\it $12^{th}$ Workshop on 
Hadron Collider Physics}, Stony Brook, NY June 5-11, 1997. For the latest 
results, see B. Klima, talk given at the {\it International Europhysics 
Conference on High Energy Physics}, Jerusalem, August 19-26, 1997.
%
\bibitem{us}
J.L. Hewett and T.G. Rizzo, hep-ph/9703337 and 9708419; 
T.G. Rizzo, hep-ph/9709235.
%
\bibitem{me}
T.G. Rizzo, in the {\it Proceedings of the 1996 DPF/DPB 
Summer Study on New Directions for High Energy Physics-Snowmass96}, Snowmass, 
CO, 25 June-12 July, 1996. 
%
\bibitem{kramer}
M. Kr\"amer, T. Plehn, M. Spira and P. Zerwas, hep-ph/9704322. 
%
\bibitem{rss}
R. R\"uckl, R. Settles and H. Spiesberger, hep-ph/9709315.
%
\bibitem{early}
J.L. Hewett and T.G. Rizzo, \PRD D36 3367 1987 . 
See also J. Bl\" umlein and R. R\" uckl,
\PLB B304 337 1993 ;
J. Bl\"umlein and E. Boos, {\it Nucl. Phys.}(Proc. Suppl.)~{\it 37B}, 
181 (1994) ;
J. Bl\" umlein, E. Boos, and A. Kryukov, \PLB B392 150 1997 ;
D. Choudhury, \PLB B346 291 1995 ;
J.E. Cieza-Montalvo and O.J. Eboli, \PRD D47 837 1993 ;
T.G. Rizzo, \PRD D44 186 1991 ;
H. Dreiner \etal, \MPL 3A 443 1988 ; G. Bhattacharyya, D. Choudhury, and
K. Sridhar, \PLB B349 118 1995 ; M.S. Berger, hep-ph/9609517.
%
\bibitem{opal}
K. Ackerstaff \etal, OPAL Collaboration, hep-ex/9708024.
%
\bibitem{egam}
J.L. Hewett and S. Pakvasa, \PLB B227 178 1989 ;
H. Nadeau and D. London, \PRD D47 3742 1993 ;
O.J. Eboli \etal, \PLB B311 147 1993 ;
G. Belanger, D. London, and H. Nadeau, \PRD D49 3140 1994 ;
F. Cuypers, \NPB B474 57 1996 ;
T.G. Rizzo, \PRD D44 186 1991 ;
M.A. Doncheski and S. Godfrey, \PRD D49 6220 1994 ~and \PRD D51, 1040  1995 .
%
\bibitem{steve}
M. Doncheski and S. Godfrey, in the {\it Proceedings of the 1996 DPF/DPB 
Summer Study on New Directions for High Energy Physics-Snowmass96}, Snowmass, 
CO, 25 June-12 July, 1996. 
%
\bibitem{GRV}
M. Gl\"uck, E. Reya and A. Vogt, Phys. Lett. {\bf B222}, 149 (1989); 
Phys. Rev. {\bf D45}, 3986 (1992); Phys. Rev. {\bf D46}, 1973 (1992).
%
\bibitem{fram}
J. Bl\"umlein and P.H. Frampton, in the {\it Proceedings of the $2^{nd}$ 
International Workshop on Physics and Experiments with Linear $e^+e^-$ 
Colliders}, Waikoloa, HI, 26-30 April 1993. See also 
P. Frampton \PRL 69 2889 1992 ~and \MPL A7 559 1992 .

\end{thebibliography}
\end{document}